\documentclass[twocolumn,showpacs,preprintnumbers,amsmath,amssymb]{revtex4}

\usepackage{graphicx}
\usepackage{color}
\usepackage{bm}
\usepackage{bbm}

\newcommand{\e}{{\rm e}}
\renewcommand{\i}{{\rm i}}

\begin{document}
\title{Wave communication across regular lattices}

\author{Birgit Hein}
\author{Gregor Tanner}
\affiliation{School of Mathematical Sciences,
University of Nottingham, University Park, Nottingham NG7 2RD, UK.}

\date{\today}

\begin{abstract}
We propose a novel way to communicate signals in the form of waves 
across a $d$ - dimensional lattice. 
The mechanism is based on quantum search algorithms 
and makes it possible to
both search for marked positions in a regular grid and to communicate between 
two (or more) points on the lattice. 
Remarkably, neither the sender nor the receiver needs to know the 
position of each other despite the fact
that the signal is only exchanged between the contributing parties. 
This is an example of using wave interference as a resource by controlling
localisation phenomena effectively.
Possible experimental realisations will be discussed. 

\end{abstract}
\pacs{03.67.Hk, 03.65Sq, 03.67.Ac, 42.50.Ex}
\maketitle

Localization phenomena in linear wave systems are closely linked to
wave interference effects. Anderson localisation in disordered media 
is a prime example thereof still posing challenges to both theory and 
experiment \cite{Hu08} 60 years after its discovery \cite{And58}. 
Recently, a new research area has emerged  focusing on 
interference as a resource and making use of 
localisation phenomena in a controlled way. Prominent 
examples are among others {\em time reversal imaging} \cite{deR} and 
reconstructing the Green function in terms of correlation functions 
\cite{Wea05}, see also \cite{TS07} . Here, information about the 
wave system is obtained by manipulating a seemingly `noisy' 
signal using phase coherence. We will focus here on another class of
wave localisation phenomena with counterintuitive properties, 
namely (quantum) search 
algorithms and (quantum) random walks. Wave search algorithms gained 
prominence with Grover's work \cite{Gro96} demonstrating a 
$\sqrt{N}$ speed-up compared to a classical search within
an unsorted data base of $N$ items. Even though search algorithms 
became an inherent part of quantum information theory 
\cite{NC00}, the speed-up is in effect caused by 
wave interference 
as has already  been pointed out by Grover \cite{GS02} and has been 
implemented for a classical wave system in \cite{BLS02}. 
Based on ideas from quantum random walks \cite{ADZ93,kempe,SJK08}, 
Grover's algorithm has been 
generalized to spatial search algorithms on networks such as on a 
hypercube \cite{SKW03, HT09} and on regular lattices \cite{AKR05}. 
Experimental realisations of quantum random walks
have  been achieved again both using classical waves (optics) 
\cite{BMK99} and quantum devices \cite{XSBL08}. 

Starting from wave search algorithms, we will demonstrate that localisation
can be used to establish communication channels across a regular 
lattice with surprising properties:
(i) signals can be exchanged exclusively  
between a source and a receiver point, 
where neither the sender nor the receiver know the 
position of each other;
(ii) the signal can track a moving receiver in the network;
(iii) the algorithm can be used as a searching device without
the necessity to know the time of measurement, (a typical requirement
for Grover's search algorithm);
(iv) the protocol can act as a sensitive switching device for 
wave transport through a lattice; 
(v) the algorithm can be effectively implemented both 
on a quantum computer and using classical waves only.
 We will first describe the set-up of the search algorithm.
We then introduce a simplified model for the search and 
explain the wave communication protocol.\\

\begin{figure}
\centering
\includegraphics[scale=0.40]{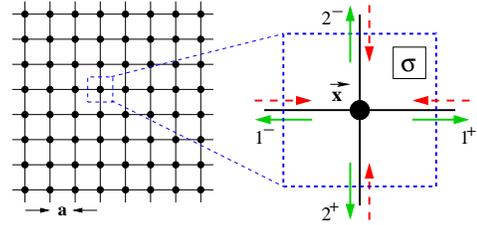}
\caption{Regular grid with $d=2, n=8$; 
local scattering within a unit cell at 
vertex $\vec{x}$ is described by the matrix $\sigma$.}
\label{fig:grid}
\end{figure}

We consider wave propagation across $d$-dimensional 
periodic lattices of identical scatterers or periodic potentials
with fixed lattice parameter $a$, see Fig.\ \ref{fig:grid}.
It is important that the lattice has a finite number of 
sites $n$ along each axis with a total
number $n^d$ of lattice sites. To simplify the calculations we will 
restrict ourselves to models with nearest-neighbour interaction only and 
consider periodic boundary conditions. 
The wave dynamics within each unit cell is given by a local
scattering matrix $\sigma$ mapping incoming channels onto outgoing 
channels, see Fig.\ \ref{fig:grid}.
($\sigma$ is also denoted a {\em coin matrix} in the context
of quantum walks). The overall wave dynamics is then given 
in terms of an operator $U_0$ 
mapping incoming onto outgoing wave coefficients between unit cells. 
We have $\mbox{dim } U_0 = 
m\, n^d$ where $m$ denotes the number of open scattering channels 
within a unit cell. Furthermore, $U_0$ is unitary when 
disregarding dissipation. Stationary solutions are obtained by the condition
\begin{equation} \label{spec} 
\det\left(1 - \e^{\i k a} U_0\right) \equiv 0
\end{equation}
where $k$ is the wave length 
and $\exp(\i ka)$ is a phase shift between incoming 
and outgoing waves. 
We neglect any (in 
general weak) $k$ dependence
of $\sigma$ and thus $U_0$.   
Note that the spectrum obtained 
from (\ref{spec}) is now periodic in $k$ with period $2 \pi/ a$; the
eigenvalues are $k_{j,l} = (2\pi l - \theta_j)/a,\, l\in \mathbb{N}$ , where 
$\theta_j$ are the eigenphases of $U_0$ with $j=1,\ldots, N$. 
To start with, we will consider a model consisting of a single open channel 
between nearby lattice sites ($m=2 d$) and we assume Kirchhoff boundary conditions 
at each vertex. 
The Hilbert space is then 
effectively $N= 2d\, n^{d}$ dimensional. This physical model captures the essence 
behind the effect described below. 

\begin{figure}
\centering
\includegraphics[scale=0.45]{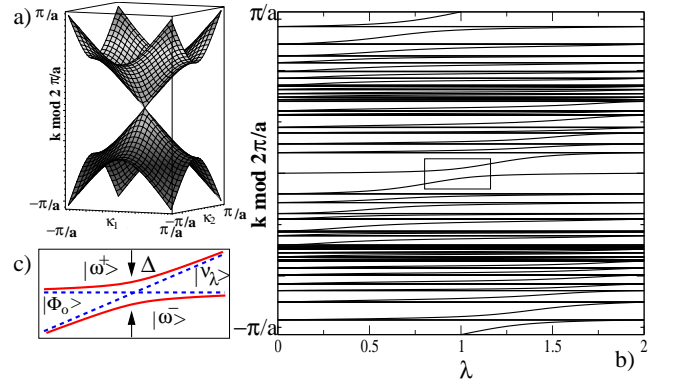}
\caption{a) The band structure at $\lambda = 0$
for $d=2$ and in the limit $n \to \infty$
with  
$\vec{\kappa}$, the Bloch wave numbers; 
b) the eigenphases of $U_{\lambda}$ for $n=11$, $d=2$; 
c) the avoided crossing with
spectral gap $\Delta$ at $\lambda=1$ and $k \mod 2\pi/a = 0$ together with 
the approximate eigenstates $|\nu_\lambda\rangle$  and 
$|\Phi_0\rangle$ (dashed lines). 
}
\label{fig:spectrum1p}
\end{figure}

We label incoming wave components from each unit cell as 
$\left| i^{\pm}\right\rangle\otimes\left| \vec{x}\right\rangle=\left| i^{\pm}, 
\vec{x}\right\rangle$, where $\vec{x}$ specifies the vertex in position 
space and $i^{\pm}$, for $i=1,\dots,d$ gives the $\pm$ direction in 
dimension $i$. Incoming waves at vertex $\vec{x}$ are mapped 
onto outgoing  waves  by a scattering matrix 
$\sigma$. For Kirchhoff boundary conditions, one obtains 
$\sigma=2\left|s\right\rangle \left\langle s\right|-\mathbbm{1}_{2d}$ and 
$\left| s\right\rangle$ is the uniform distribution 
$\left| s\right\rangle=\frac{1}{\sqrt{2d}}
\sum_{i=1}^{d} \left(\left|i^{+}\right\rangle +
\left|i^{-}\right\rangle\right)$ \cite{AKR05}. 
Outgoing waves in direction $i^{\pm}$ are now
identified with incoming waves at an adjacent vertex
$\vec{x}\pm a \, \vec{e}_i$ where $\vec{e}_i$ is the unit vector
in direction $i$. The local scattering processes is described 
in terms of a (global) scattering (or coin) matrix 
$C=\sigma \otimes \mathbbm{1}_{n^{d}}$. The full wave propagator 
(or quantum walk) $U_0$ is obtained from $C$ after identifying 
incoming and outgoing waves of adjacent unit cells accordingly. 
The spectrum of the unperturbed walk $U_0$ exhibits a band structure 
as shown in Fig.\ \ref{fig:spectrum1p} a), here for $d=2$.
(For a finite lattice, the quasi-momenta $\kappa_x, \kappa_y$ are discretised 
according to $\kappa_{x,y} = 2 \pi j/ (n a); \, j=0 \ldots n-1$.)\\

Following Ambainis, Kempe and Rivosh (AKR), 
the quantum walk $U_0$ acts as a search algorithm after marking a target 
vertex $\left| v\right\rangle$ by a modified scattering matrix 
$\sigma'$ \cite{AKR05}, that is, one considers
$C^{\prime}=C-\left(\sigma-\sigma'\right)\otimes \left|v\right\rangle 
\left\langle v\right|$.  The AKR search uses $\sigma'=-\mathbbm{1}_{2d}$.  
Since $\left|s\right\rangle$ is 
an eigenvector of $\sigma$, we may write $U^{\prime}=
U_0 \left(1 -2\left|sv\right\rangle\left\langle sv\right|\right)$, 
where $\left|sv\right\rangle=\left|s\right\rangle\otimes\left|
v\right\rangle$. The search algorithm is initialised in the uniform state 
$\left|\Phi_0 \right \rangle = 1/\sqrt{N} 
\sum_{\vec{x}} \left|s\vec{x}\right \rangle$, and the walk 
$(U')^T\left|\Phi_0\right\rangle$ localizes at $|{v}\rangle$ after 
$T \propto \sqrt{N}$ steps.

The AKR search can be 
analysed 
by defining a one parameter family of unitary operators \cite{HT09}
\begin{equation} \label{qwalk}
U_{\lambda}=
U_0+\left(\e^{\i\pi\lambda}-1\right) 
U_0\left|sv\right\rangle\left\langle sv\right|;
\end{equation}
one obtains $U_0$ for $\lambda = 0$ or 2 and the 
AKR search for $\lambda = 1$. The part of the
eigenfrequency spectrum of $U_\lambda$ interacting with 
the perturbation is shown in Fig.~\ref{fig:spectrum1p} b). 
The spectrum is periodic in $k$ with 
period $2\pi/a$ independent of $\lambda$. When varying $\lambda$, 
a ``perturber state'' $|\nu_\lambda\rangle$
emerges which crosses the $k \mod 2\pi/a =0$ axis at $\lambda = 1$. 
The resulting avoided crossing between the initial state 
$|\Phi_0\rangle$ and $|\nu_\lambda\rangle$ 
is shown in Fig.\ \ref{fig:spectrum1p} c). Note that $|\Phi_0\rangle$ is
the fully symmetric eigenstate of $U_0$ corresponding 
to a $d$-dimensional Bloch-vector $\vec{\kappa} =0$ of the 
unperturbed spectrum with eigenvalue $k\mod 2\pi/a = 0$. 
Like in Grover's algorithm, the quantum search $U' = U_{\lambda=1}$ 
rotates the initial state $|\Phi_0\rangle$ into a localised state 
$|\nu_\lambda\rangle$ which has here a strong overlap with
the target state $|sv\rangle$. The search time $T_0$
is inversely proportional to the gap at the avoided crossing 
$\Delta$, that is, $T_0\approx {\pi}/{\Delta}$.

In order to obtain an estimate for the search time $T_0$ as well as the 
efficiency of the search, that is, the matrix element 
$\langle sv|\nu_\lambda \rangle$, it is essential to find the
approximately invariant two-level subspace near the crossing 
spanned by $|\Phi_0\rangle$ and $|\nu_{\lambda=1}\rangle$.
The technique developed in \cite{HT09} for the hypercube 
has been adapted to regular grids. We will only give the  result 
here, further details will be presented elsewhere \cite{HT09I}.
One finds for the normalised vector $\left|\nu_{\lambda=1}\right\rangle$
\begin{eqnarray} \label{t-state}
\left|\nu_{\lambda=1}\right\rangle
&=&- \langle sv|v_{\lambda=1}\rangle  \sqrt{\frac{2}{N}} 
\sum_{\vec{\kappa}\neq\vec{0}} 
\e^{-\frac{2 \pi \i}{n} \vec{\kappa}\vec{v}}\times\\\nonumber
&&\left(\frac{\e^{\i\theta_{\vec{\kappa}}}}{1 - \e^{\i\theta_{\vec{\kappa}}}}
\left|\Phi_{\vec{\kappa}}^{+}\right\rangle
+\frac{\e^{-\i\theta_{\vec{\kappa}}}}{1 -\e^{-\i\theta_{\vec{\kappa}}}}
\left|\Phi_{\vec{\kappa}}^{-}\right\rangle\right),
\end{eqnarray}
where $\vec{v}$ is the position of the target vertex and 
$|\Phi^{\pm}_{\vec{\kappa}}\rangle$ and $\pm\theta_{\vec{\kappa}}$ 
are the eigenvectors and eigenphases of the unperturbed 
walk $U_0$ \cite{AKR05}. 
The $d$-dimensional label $\vec{\kappa}$ with $\kappa_i = 0,\ldots, n-1$
is equivalent to the (discretised) 
Bloch wave number, see Fig.\
\ref{fig:spectrum1p} a). The eigenphases are explicitly given as 
$\cos\theta_{\vec{\kappa}}= \frac{1}{d}\sum_{i=1}^{d}
\cos\frac{2\pi \kappa_{i}}{n}$. The overlap-matrix element 
$\langle sv|\nu_{\lambda=1}\rangle$ can be estimated 
\cite{HT09I}:
\[
\langle sv|\nu_{\lambda=1}\rangle = \left\{ 
  \begin{array}{lcl} {\cal O}(1/\sqrt{\log N)} &\mbox{for}& d =2,\\ 
                     {\cal O}(1) &\mbox{for}& d > 2 .
  \end{array} \right. 
\]
Detailed expressions for the leading order coefficients for $d=2$ and 3 are
given in \cite{HT09I}.  We find  that $|\nu_{\lambda=1}\rangle$ is 
exponentially localised on the marked vertex $\vec{v}$;
note, that the overlap of $|sv\rangle$ with 
a typical eigenstate of the unperturbed spectrum $|\Phi_\kappa\rangle$ is of 
the order ${\cal O}(N^{-\frac{1}{2}}) \ll {\cal O}(1)$. 

Near the avoided crossing, the level dynamics can be described in terms of
the two-level sub-space spanned by the orthogonal vectors 
$|\nu_\lambda\rangle$ and $|\Phi_0\rangle$. Writing the unitary operator 
$U_{\lambda}$ as $U_{\lambda}=\e^{-\i H_{\lambda}}$, one obtains at 
$\lambda =1$ in the $\{|\Phi_0\rangle, |\nu_{\lambda=1}\rangle\}$ basis 
an effective two-dimensional Hamiltonian $H_{\lambda = 1}$ 
of the form 
\begin{equation} \label{2x2H}
H^{2\times 2}=\begin{pmatrix}  k_l a  & - \i\epsilon \\ 
\i\epsilon & k_l a 
\end{pmatrix}
\end{equation}
with $k_l = 2 \pi l/a, l\in \mathbb{N}$ being the 
wave numbers corresponding to $\vec{\kappa} = 0$ 
states of the 
unperturbed lattice and $\epsilon$ is a real and 
positive coupling parameter, that is, 
\begin{equation} \label{eps} 
\epsilon = \frac{\Delta}{2} =
\langle \nu_{\lambda=1} |U_1| \Phi_0 \rangle  \approx
\frac{2\left|\langle sv \mid \nu_{\lambda}\rangle \right|}{\sqrt{N}}+{\cal O}\left(N^{-1}\right)
\end{equation}
with $\Delta$, the gap at the avoided crossing.  

\begin{figure}
\centering
\includegraphics[scale=0.5]{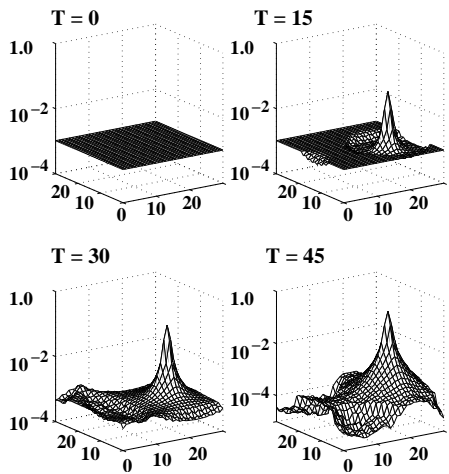}
\caption{Probability distribution of the quantum walk on a 
$31\times 31$-grid 
up to $T=45$ time steps.} 
\label{fig:search}
\end{figure}

The start vector $(1,0) \equiv |\Phi_0\rangle$ is rotated into the 
localised state $(0,1) \equiv |\nu_{\lambda=1}\rangle$ in $T_0 = \pi/(2\epsilon)$ 
steps leading to the $\sqrt{N}$ speed-up \cite{AKR05}.  
The whole process is  $2 \,T_0$-periodic , that is, one needs 
- like for Grover's algorithm \cite{Gro96} - to 
know the period $T_0$ to perform the search. 
For a simulation of the search on a $31\times 31$ grid, see
Fig.\ \ref{fig:search}.\\

Interesting applications emerge when considering several 
target vertices, $\left| v^{i}\right\rangle, i=1,\ldots m$ with $m\ll N$. 
We now define a set of parameters $\lambda=(\lambda_1, \ldots, \lambda_m)$ 
and a search  algorithm of the form
\begin{equation}
U_{\lambda}=U+\sum_{i=1}^{m}\left(\e^{\i\pi
\lambda_{i}}-1\right) U\left|sv^{i}\right\rangle\left\langle sv^{i}
\right| .
\end{equation}
At 
$\lambda=\left(1,1,\dots,1\right)$ and $ka\mod 2\pi = 0$, one 
finds that there are $m-1$ degenerate eigenvalues and two
further eigenvalues forming an avoided crossing with the degenerate
subset. 
The corresponding set of $m+1$ eigenstates coincides in good
approximation with the subspace spanned by the uniform distribution 
$|\Phi_0\rangle$ and now $m$ localised states 
$\left|\nu_{\lambda}^{i}\right\rangle, i=1,\ldots,m$. 
Each of the $\left|\nu_{\lambda}^{i}\right\rangle$
is well described by the approximation (\ref{t-state})
and $\langle \nu_{\lambda}^{i}|\nu_{\lambda}^{j}\rangle \approx \delta_{ij}$.
The localised states interact at the crossing predominantly 
via $|\Phi_0\rangle$ which takes on the role of a {\em carrier state}.
In analogy to (\ref{2x2H}), we write a model Hamiltonian at the crossing  
in the basis
$\{|\Phi_0\rangle, |\nu_{\lambda=1}^{1}\rangle, \ldots, 
|\nu_{\lambda=1}^{m}\rangle \}$  as
\begin{equation}
H^{\left(m+1\right) \times \left(m+1\right)}=
\begin{pmatrix} 
k_l a            & - \i\epsilon & - \i\epsilon & \dots & - \i\epsilon \\
\i\epsilon   &  k_la   & 0           & \dots    & 0  \\
\i\epsilon   &     0      &  k_l a    &  \ddots & \vdots \\
\vdots                   & \vdots   &  \ddots  & \ddots  &0 \\
\i\epsilon  &     0      &  \dots    &  0        & k_la  \\
\end{pmatrix}
.\end{equation}

Like for the full propagator $U_{{\lambda} = 1}$, the spectrum of 
$H^{\left(m+1\right) \times \left(m+1\right)}$ consists of
$\left(m-1\right)$ eigenvalues equal to $k_l a$ and two eigenvalues 
$k_l a \pm \sqrt{m} \epsilon$ with eigenvectors 
$|\omega^{\pm} \rangle= \i/\sqrt{2 m}
\left(\mp \i \sqrt{m},1,\dots,1\right)^{\rm t}$. 
The gap between the non-degenerate levels is now $\Delta = 2 \sqrt{m} 
\epsilon$ with $\epsilon$ given in (\ref{eps}).
Starting the search in the totally symmetric 
state $|\Phi_0\rangle$ at $\lambda = (1,\ldots,1)$, one finds 
that $U^T_{\lambda}|\Phi_0\rangle$ localises 
on \underline{all} $m$ marked vertices after 
$T_0 \sim \pi/2 \sqrt{N/m}$ (or, for $d=2$, $T_0 \sim \pi/2 \sqrt{N/m\log N}$) 
steps simultaneously. \\

More interestingly, the quantum walk can also be used to transmit signals
across the network. Such a sender-receiver configuration has to the best of 
our knowledge not been described before and may have interesting applications 
both in a quantum setting, but also for classical waves (such as microwaves,
in optics or acoustics).  Instead of starting the 
walk in the uniform state $|\Phi_0\rangle$, we propose to begin the walk
at one of the localised states 
$|\nu_{\lambda}^{m}\rangle$, say. At the avoided
crossing, this state can approximately be described as
\begin{equation}
|\nu_{\lambda}^{m}\rangle = - \frac{\i}{\sqrt{2m}} \left( |\omega^+\rangle + |\omega^-\rangle
- \i \sqrt{2(m-1)} |\omega_0\rangle \right)
\end{equation}
with 
$|\omega_0\rangle = (m(m-1))^{-1/2} (0,1\ldots,1-m)^t$ in the basis 
spanning 
$H^{(m+1)\times(m+1)}$; $|\omega_0\rangle$ is a vector in the degenerate 
eigenspace with eigenvalue $k_l a$. Applying the walk for 
$T_s= \pi/(\epsilon\sqrt{m}) = 2 T_0$ steps leads to
\begin{eqnarray} \label{sender}
U^{T_s}|\nu_{\lambda}^{m}\rangle &=&  
\frac{\i\e^{\i k_l a T_s}}{\sqrt{2 m}} \left( |\omega^+\rangle + |\omega^-\rangle
+ \i \sqrt{2(m-1)} |\omega_0\rangle\right) \\\nonumber
&=& \e^{\i k_l a T_s} 
\left(0,-\frac{2}{m},\ldots,-\frac{2}{m},1-\frac{2}{m}\right)^t .
\end{eqnarray}
This implies that a signal of intensity $4/m^2$  is transmitted from the 
sender (located at the $m$-th marked vertex) to each of the $m-1$ other 
marked vertices. The intensity at the sender at time $T_s$ is then of 
the order $(1-2/m)^2$. These findings have been verified numerically in our 
model.
Interestingly, in the case $m=2$ with a single receiver, the 
signal is transmitted in \underline{full}. This opens up the possibility  
of transferring signals directly between two points on a network where 
neither the sender nor the receiver know each other's position. In 
addition, the sender has information about the number of 
receivers by recording the signal at time $T_s$. 

\begin{figure}
\centering
\includegraphics[scale=0.30, angle=-90]{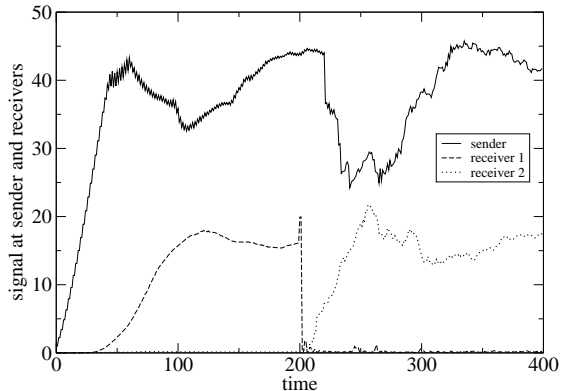}
\caption{Sending continuously: signal at sender (full) and
receivers (dashed/dotted). The response of the system to switching 
the position of the receiver at $T=200$ is shown.
(We introduced a damping of 1\% at all vertices.) 
}
\label{fig:continuous}
\end{figure}

The effect persists also for a continuous source at the sender position.
In Fig.\ \ref{fig:continuous}, we recorded the signal both
at the sender and at the receiver. (To keep the signal finite, 
we added a small amount of absorption across the network). Again, 
without prior knowledge of the receiver's 
position, the network localises at the two marked vertices, thus making
it possible to actually exchange information continuously between these
two points. Changing the position of the receiver leads to
a sudden drop of the signal at the old receiver position and a build-up 
at the new position, see Fig.\ \ref{fig:continuous}. The system is thus
capable of tracking a moving receiver position! The signal speed is 
limited by the transfer time $T_s \sim \sqrt{N}$ ($T_s \sim \sqrt{N \ln N}$ 
for d=2) and thus is the speed at which the receiver can move.

The continuous sender/receiver protocol can also be used to search 
a marked item without knowing the search time $T_0$; 
this is a serious complication of Grover-type search algorithms when the 
precise number of marked items is not know (as $T_0$ depends on $m$). Here, 
we find the marked items as long as we wait for times $T\ge T_s$.
Furthermore, the system can act as a switching device. Wave transport 
between two points on the grid can only be achieved, if the system
is tuned to the avoided crossing. Slight detuning by for example changing the 
parameter $\lambda$ in (\ref{qwalk}) will quickly cut-off the signal.\\

The described effects open up completely new ways of transmitting signals
across regular networks. While a construction of the map $U$ is certainly 
feasible in a quantum setting and can be implemented efficiently on a 
quantum computer \cite{NC00}, an implementation using classical waves 
may be even more promising. For dispersionless wave dynamics and on regular
lattices, the unitary matrix $U_{\lambda}$ is equivalent to a 
discretised version of the time dependent Green function (for times 
$t= a/c$ with $c$ being the wave velocity; corresponding time scales in 
a quantum setting would be given by the group velocity \cite{KP09}). Indeed, 
localised states due to a local perturbation in 
a regular lattice are well known; (for example in optical crystals, see
\cite{Joa08}). We predict that the effect will occur if ``defect states'' 
created by local phase perturbations (equivalent to the perturbed coin $C'$) are
pushed into the (discretised continuum) of the band close to the fully periodic
state - the $\vec{\kappa} =0$ state, see Fig.\ \ref{fig:spectrum1p} a. 
Furthermore, the interaction between the defect states and the lattice states 
must be small enough to lead to avoided crossings between these two
states only. We expect that a signal (such as a laser or a 
microwave transmitter coupled into a periodic structure at a defect position) 
can be transmitted \underline{and} focused onto another 
defect in the same lattice using the described effect. The totally symmetric 
state $|\Phi_0\rangle$ acts then as a carrier state guiding the signal
between the perturbations.\\

We thank J Keating, U Kuhl, T Monteiro, U Peschel, H-J St\"ockmann 
and H Susanto for helpful discussions and S Gnutzmann for
carefully reading the manuscript.


\begin{thebibliography}{1}
\bibitem{Hu08} H.\ Hu {\em et al}, {\em Nature Physics} {\bf 4}, 945 (2008)
and ref.\ therin.
\bibitem{And58} A.\ Anderson, {\em Phys.\ Rev.} {\bf 109}, 1492 (1958).

\bibitem{deR} J.\ de Rosny and M.\ Fink, {\it Phys.\ Rev.\ Lett.} 
{\bf 89}, 124301 (2002). 


\bibitem{Wea05} R.\ L.\  Weaver, {\em Sience} {\bf 307}, 1568 (2005).

\bibitem{TS07} G.\  Tanner and N.\ S\o ndergaard, {\em J.\ Phys.\ A} {\bf 40}, 
R443 (2007).


\bibitem{Gro96} L.\ \ K.\ Grover, 
{\it Phys.\ Rev.\ Lett.}, {\bf 79} 325 (1997).

\bibitem{NC00} M.\ A.\ Nielsen and I.\ L.\ Chuang, 
{\it Quantum Computation and Quantum 
Information}, (Camb.\ Uni.\ Press 2000).

\bibitem{GS02}
L.\ K.\ Grover and A.\ M.\ Sengupta, 
{\it Phys.\ Rev.\ A} {\bf 65}, 032319 (2002).

\bibitem{BLS02} N.\ Bhattacharya, H.\ B.\ Linden van den Heuvell and 
R.\ J.\ C.\ Spreeuw, {\it Phys.\ Rev.\ Lett.\ } {\bf 88} 137901 (2002).

\bibitem{ADZ93} Y.\ Aharonov, L.\ Davidovich and N.\ Zagury,
{\em Phys.\ Rev.\ A} {\bf 48}, 1687 (1993).

\bibitem{kempe} J.\ Kempe, 
{\it Contemporary Physics}, {\bf 44} 307 (2003)

\bibitem{SJK08} M.\ Stefanak {\em et al}, 
{\it Phys.\ Rev.\ Lett.\ } {\bf 100}, 020501 (2008).

\bibitem{SKW03} N.\ Shenvi, J.\ Kempe and K.\ B.\ Whaley, 
{\it Physical Review A}, {\bf 67} 052307 (2003).

\bibitem{HT09} B.\ Hein and G.\ Tanner, {\it J. Phys. A} 
{\bf 42} 085303 (2009).

\bibitem{AKR05} A.\ Ambainis, J.\ Kempe and A.\ Rivosh, 
{\it Proceedings of the  16th ACM-SIAM SODA} (SIAM, Philadelphia, 2005),
 p.\ 1099.

\bibitem{BMK99} 
D.\ Bouwmeester {\em et al}, 
{\em Phys.\ Rev.\ A} {\bf 61}, 013410 (1999);
P.\ L.\ Knight {\em et al}, 
{\em Phys.\ Rev.\ A} {\bf 68} 020301 (R) (2003).

\bibitem{XSBL08} P.\ Xue {\em et al}, 
{\em Phys.\ Rev.\ A} {\bf 78}, 042334 (2008);
H.\ Schmitz {\em et al}, 
{\em Phys.\ Rev.\ Lett.} {\bf 103}, 090504.

\bibitem{HT09I} B.\ Hein and G.\ Tanner, in preparation.

\bibitem{Joa08} J.\ D.\ Joannopoulos {\em et al}, {\em Photonic Crystals}, 
(Princ.\ Uni.\ Press, Princeton, 2008).

\bibitem{KP09} A.\ Kempf and R.\ Portugal, 
{\em Phys.\ Rev.\ A} {\bf 79}, 052317 (2009).

\end{thebibliography}
\end{document}